\documentclass[aps,twocolumn,showpacs,floatfix,amssymb]{revtex4}
\usepackage{graphicx}
\begin{document}

\title{Superconductivity of strongly correlated electrons
on the honeycomb lattice}

\author{ A.A. Vladimirov$^{a}$, D. Ihle$^{b}$ and  N. M. Plakida$^{a,c}$ }
 \affiliation{ $^a$Joint Institute for Nuclear Research,
141980 Dubna, Russia}
 \affiliation{$^{b}$ Institut f\"ur Theoretische Physik,
 Universit\"at Leipzig,  D-04109, Leipzig, Germany }
\affiliation{$^{c}$ Max-Planck-Institut f\"ur Physik Komplexer Systeme D-01187,
Dresden, Germany}
\thanks{\emph{ E-mail:} plakida@theor.jinr.ru
}%

\date{\today}

\begin{abstract}
A microscopic theory of the electronic  spectrum and of superconductivity
 within the $t$-$J$ model on the honeycomb lattice is developed. We derive the
equations for the normal and anomalous Green functions  in terms of the Hubbard operators
by applying the  projection technique.   Superconducting pairing of  $d+id'$-type
mediated by the antiferromagnetic exchange  is found. The superconducting $T_c$ as a
function of hole doping exhibits a two-peak structure related to the van Hove
singularities of the density of states for the two-band $t$-$J$ model. At half-filling
and for large enough values of the exchange coupling, gapless superconductivity may
occur. For small doping the coexistence of antiferromagnetic order and superconductivity
is suggested. It is shown that the $s$-wave pairing is prohibited, since it violates the
constraint  of no-double-occupancy.
\end{abstract}



\maketitle

\section{Introduction}
\label{sec:1}

Graphene, the two-dimensional carbon honeycomb lattice,  has been recently extensively
studied due its peculiar electronic properties caused by the low energy  cone-type
electronic spectrum at the Dirac points $K$ and $K'$ in the Brillouin zone (BZ) (for a
review see~\cite{Neto09}). The effects of electronic interactions play a minor role close
to half-filling at low density of electronic states at the Dirac points, but for large
doping   interactions appear to be important. Studies of graphene beyond the simple model
of noninteracting electrons  by taking into account the Coulomb interaction  reveal a
rich phase diagram with phase transitions to the antiferromagnetic (AF) state,
spin-density wave (SDW), charge-density wave (CDW), and  unconventional superconductivity
(SC)  (for a review see~\cite{Kotov12}).

The superconducting order parameters in the two-dimensional honeycomb lattice described
by the hexagonal symmetry group $D_{6h}$ have a complex character. A general symmetry
analysis of available irreducible representations (IR) and  superconducting order
parameters is given in Ref.~\cite{Black14}. In the case of the singlet pairing, the
extended  $s$-wave $s_{x^2 + y^2}$ order parameter   ($E_{1g}$ IR), $d_{x^2 - y^2}$ and
$d_{xy}$ order parameters  ($E_{2g}$ IR) preserving the time-reversal symmetry and its
time-reversal symmetry breaking complex combination $d_{x^2 - y^2}\pm id_{xy}\;  (d \pm
id')$ are commonly discussed. The triplet pairing with the $p_x +  ip_y\; (p+ip')$ order
parameter ($E_{1u}$ IR) is also often discussed in the literature. This symmetry
consideration is very important in discussing SC in graphene. But it can be  also applied
in studies of other electronic systems with  the two-dimensional honeycomb lattice, such
as the transition metal dichalcogenides~\cite{Uchoa05}.

Several models of the Bardeen-Cooper-Schrieffer-type (BCS) were discussed. In
Ref.~\cite{Uchoa07} a model with the on-site and the nearest-neighbor (nn)  electron
interactions of the BCS-type was considered. Assuming bond-independent anomalous
correlation functions on the nn lattice sites, the $s$-wave  pairing  with ${\bf
k}$-independent  gap induced by the on-site interaction and the extended $s$-wave pairing
induced by the nn interaction were found.   At the Dirac points close to half filling the
latter can be described as $p+ip$ phase.  At large value of coupling constants a gapless
SC emerges at half-filling.

The superconducting phase transition caused by the Coulomb interaction was studied within
the Hubbard model\cite{Hubbard63} on the  honeycomb lattice in a wide range of the
on-site Coulomb repulsion $U/t $  from weak to strong coupling (in graphene  $U/t = 4 -
5$ and $t\sim 2.8$~eV~\cite{Neto09}).  The renormalization group (RG) approach was used
in~\cite{Honerkamp08} to study phase transitions in the extended Hubbard model with the
on-site interaction $U$, the nn  repulsion $V$, and the spin-exchange interaction $J$.
Close to half-filling, the SDW or CDW orders occur for large $U$ and $V$, while  for a
large doping  $f$-wave triplet-pairing and $d+id'$-wave singlet-pairing emerge. In
Ref.~\cite{Faye15} the extended Hubbard model for graphene  with the nn  repulsion $V $
and on-site interaction $U $  was considered. Using the variational cluster approximation
and the cellular dynamical mean-field theory  the  SC of different  symmetries was
studied.  Depending on the values of $ V $ and $U$, the triplet $p$-wave symmetry and the
chiral combination $p + ip'$ were found, while the singlet SC (extended $s$- or $d$-wave)
was not clearly detected.

Superconductivity on the  honeycomb lattice  is commonly  studied also within the
phenomenological $t$--$J$ model, where the exchange interaction ($J \sim t$) is
considered as a fitting parameter. In Ref.~\cite{Black07} SC was studied in a graphite
layer within the resonating valence bonds (RVB) approach~\cite{Baskaran02} for the
$t$--$J$ model.  Both  the extended $s$-wave and $d$-wave pairing  with the order
parameter determined by bond-dependent anomalous correlation functions  were considered.
The superconducting $T_c$ for the $d$-wave pairing appears to be much larger than for the
extended $s$-wave pairing and has a high value with a  maximum at doping $\delta = 0.25$.
A similar model was considered in Ref.~\cite{Jiang08} for the $d +id'$-pairing with the
bond-dependent anomalous correlation functions. The spectrum of electronic excitations in
the superconducting state is determined by two gaps $g_{\pm}= |\Delta(\pm {\bf k})|$ with
spin up in one  valley and spin down in the other valley with zeros at $K$ and $K'$
points of the BZ, respectively. The excitations are gapless at half filling for any value
of the coupling constant.  The $t$--$J$ model  with the on-site interaction of
intermediate strength $U/t = 2.4$ was considered in Ref.~\cite{Pathak10} using the
variational Monte Carlo study.  The superconducting ground state with the $d+id'$ pairing
for doping $0 < \delta < 0.4$  was found. It was estimated that the superconducting $T_c$
can reach  room temperatures at an optimal doping around $\delta = 0.15$. The extended
$t$--$J$ model with the nn and the next-nearest-neighbor (nnn) exchange interaction $J_1$
and $J_2$,  respectively,   was considered in Ref.~\cite{Wu13}.  In the heavily doped
case (around 3/8 and 5/8 filling), a chiral $d + id'$ symmetry was obtained. The
competition between antiferromagnetism and SC in the vicinity of half filling was
considered by applying  the functional RG.

The quantum phase diagrams of both the Hubbard model and the $t$--$J$ model on the
honeycomb lattice at 1/4 doping were studied in Ref.~\cite{Jiang14}. At this doping, in
the nn tight-binding model the nested Fermi surface emerges which is unstable in the
presence of a weak interaction. Using a combination of exact diagonalization, density
matrix renormalization group, the variational Monte Carlo method, and quantum field
theories, it was shown that in a wide range of the Hubbard repulsion, $1 < U/t < 40$, or
the exchange interaction, $0.1 < J/t < 0.8$,  the quantum ground state is either a chiral
SDW state or a spin-charge-Chern liquid, but not a $d+id'$ superconductor. For the
$t$--$J$ model  at larger $J/t > 0.8$ a first-order phase transition to the $d+id'$
superconductor occurs.

A detailed study of the $t$--$J$ model on the honeycomb lattice was presented in
Ref.~\cite{Gu13}. Using the Grassmann tensor product state approach, exact
diagonalization and density-matrix renormalization methods, the ground-state energy,  the
staggered magnetization in the AF phase, and the SC order parameter  as a function of
doping $\delta$  have been calculated. The occurrence of the time-reversal symmetry
breaking $d + id'$-wave SC up to $ \delta =0.4$  was found. Moreover, a coexisting of the
SC and AF order was observed for  low doping, $ 0 <\delta< 0.1$, where the triplet
pairing is induced (see also~\cite{Gu14}). In Ref.~\cite{Black14}  SC on the honeycomb
lattice close to the Mott state at half filling was studied within  the $t$--$J$  model
using the renormalized mean-field theory  and in the Hubbard model by  quantum Monte Carlo
calculations. It was shown  that the chiral $d + id'$-wave state is the most favorable
state for  a wide range of the on-site repulsion $U$. At the same time, a mixed chirality
$d$-wave state, such as a state with $d + id'$-wave symmetry in one valley but $d -
id'$-wave symmetry in the other valley, is not possible in the $t$--$J$ model without
reducing the translational symmetry. No energetically favorable $d$-wave solution with an
overall zero chirality  was found within the  $t$--$J$  model.

The van Hove singularity (VHS) scenario of SC being developed for cuprates (see, e.g.,
~\cite{Markiewicz97}) was discussed  in several publications.  In Ref.~\cite{McChesney10}
the extended VHS in doped   graphene  was found using the angle-resolved photoemission
spectroscopy. Considering the conventional fluctuation exchange
approximation~\cite{Scalapino87} with the weak Hubbard interaction  $U/t \lesssim 4 $,
the competition between  magnetic instability and  SC was analyzed. It was found that  SC
plays a dominant role when the Fermi level is placed close enough to the extended VHS,
where the transition temperature $T_c$  can be quite high. In Ref.~\cite{Gonzalez08} it
was shown that, due to the strong anisotropy of the electron scattering at the VHS,
attractive coupling channels appear  from the  originally repulsive interaction that
results in the  superconducting pairing with  $T_c  \sim 10$~K. In Ref.~\cite{Ma14}
studies of the Hubbard model on the honeycomb lattice  with nn and nnn interactions show
the appearance of the extended VHS, where the density of states diverges in a power law.
Using the random-phase-approximation and determinant quantum Monte Carlo approaches  a
possible triplet $p + ip'$ SC with relatively high $T_c$  was found at low filling. The
interplay between SC and SDW order in graphene close to the VHS was considered  in
Ref.~\cite{Nandkishore12}. The instabilities  to both the chiral $d + id'$ SC and  the
uniaxial SDW were found in a model with four different interactions between fermions near
saddle points. The SC  is strongest at the VHS, but slightly away from it  SDW  appears
first. To investigate the possibility of coexistence of SC and SDW,  the Landau-Ginzburg
functional was derived. It was shown that SDW does not coexist with SC, because both
phases  are separated by  first-order transitions. The dynamic cluster approximation was
used in Ref~\cite{Xu16} to study SC in the Hubbard model with $U/t = 2 - 6$.  A
transition from the $d + id'$-wave singlet pairing, which dominates  close to the VHS
filling, to the $p$-wave triplet pairing at larger coupling was found.

In several studies the renormalized mean field theory for the $t$--$J$ model was
employed. To take into account strong correlations of electrons in the singly occupied
band,  the hopping parameter $t$ and exchange interaction $J$ were renormalized by the
statistical weighting factors $g_t = 2\delta/ (1+\delta)$ and $g_J = 4/ (1+\delta)^2$, as
in the RVB theory for cuprates~\cite{Zhang88}. We point out that this renormalization is
rather crude, e.g., for $\delta = 0$ it results in the zero band-width $\sim \delta t$
though at low doping spin-polaron quasiparticles  appear with a finite bandwidth of the
order $J$ (see, e.g.~\cite{Martinez91}). Moreover, in the undoped case the four times
stronger exchange interaction $4 J$ results  while  the Heisenberg model with the
original  exchange interaction $J$ should be found. The slave-boson approximation also
strongly violates the statistics of the projected electrons in the original $t$--$J$
model~\cite{Anderson87}. To take into account the restriction of no-double-occupancy in
the $t$--$J$ model, a technique for the projected electron operators,  the Hubbard
operators~\cite{Hubbard65}, should be used.

In the present paper we develop a microscopic theory of SC of strongly correlated
electrons on the honeycomb lattice employing the Hubbard operator technique. This
technique  was used in our previous paper for the calculation of the electronic spectrum,
the spin-excitation spectrum, and of thermodynamic quantities  within the $t$--$J$
model~\cite{Vladimirov18}. Using the projection operator technique~\cite{Mori65}
developed for the thermodynamic Green functions (GFs)~\cite{Zubarev60}  in
Ref.~\cite{Plakida11},  we derive equations for the normal and anomalous (pair) GFs. In
the generalized mean-field approximation (GMFA) a self-consistent system of equations for
the singlet order parameters  is obtained and the superconducting $T_c$ as a function of
doping $\delta$ is calculated. It is shown that the condition of the no-double-occupancy
of quantum states in the $t$--$J$ model is violated for the $s$-wave pairing, while  the
$d+id'$ pairing obeys this restriction.

The paper is organized as following. In  Section~\ref{sec:2} the $t$--$J$ model for the
honeycomb lattice is formulated. Equations for the GFs are derived in
Section~\ref{sec:3}. Gap equations and the calculation of $T_c$ are given in
Section~\ref{sec:4}. The conclusion is presented in Section~\ref{sec:5}.

\section{The $t$-$J$ model }
\label{sec:2}

The Hubbard model~\cite{Hubbard63}  is commonly used in  studies of  correlated
electronic systems. In the limit of strong correlations the model  is reduced to the
one-subband  $t$-$J$  model~\cite{Anderson87}. In the lattice site representation the
model reads:
\begin{eqnarray}
H & = & - t\, \sum_{\langle i,j \rangle \sigma}  \tilde a_{i,\sigma}^{+} \tilde
a_{j,\sigma} - \mu \sum_{i , \sigma} \,n_{i,\sigma}
\nonumber\\
&+& \frac{J}{2}\sum_{\langle i, j \rangle}\;\left( {\bf S}_{i} \; {\bf S}_{j}
 - \frac{1}{4} n_i\, n_j \right),
  \label{1}
\end{eqnarray}
where $\tilde a_{i,\sigma}^{+} =a_{i,\sigma}^{+}(1-n_{i ,\bar{\sigma}})\;$ and $ \tilde
a_{i\sigma} =a_{i\sigma}(1-n_{i ,\bar{\sigma}})$ are projected creation and annihilation
electron operators on the site $i$ with spin $\sigma/2$ ($\sigma=\pm 1 , \; \bar{\sigma}
= - \sigma$), and the number operator $n_{i}  = \sum_{ \sigma}\tilde a_{i,\sigma}^{+}\,
\tilde a_{i,\sigma}$.   Here $\langle i, j \rangle$ denote the nearest
neighbors  for electron hopping with energy $\,t\,$  and for spins ${\bf S}_{i}$ with  AF
exchange interaction $J$.

It is convenient to employ the Hubbard operator (HO) technique~\cite{Hubbard65} where the
projected  electron operators are written as:  $\tilde a_{i\sigma}^{+} = X_{i}^{\sigma 0}
\; $,  $\tilde a_{j\sigma}= X_{j}^{0\sigma}$. The HOs  $\, X_{i}^{nm}=|i,n\rangle\langle
i,m| \,$  describe transitions between  three possible states at a lattice site $i$:
$|i,n\rangle=|i,0\rangle$  and $|i,\sigma\rangle$ for an empty site and for a singly
occupied site by an electron with spin $\sigma/2$, respectively.

The electron number operator and the spin operators  are defined as
\begin{eqnarray}
  n_i &=& \sum_{\sigma} X_{i}^{\sigma \sigma} =   X_{i}^{++}  +  X_{i}^{--},
\label{3a}\\
S_{i}^{\sigma} & = & X_{i}^{\sigma\bar\sigma} ,\quad
 S_{i}^{z} =  (\sigma/2) \,[ X_{i}^{\sigma \sigma}  -
  X_{i}^{\bar\sigma \bar\sigma}] .
\label{3b}
\end{eqnarray}
The commutation relations for the HOs read:
\begin{equation}
\left[X_{i}^{nm}, X_{j}^{kl}\right]_{\pm}= \delta_{ij}\left(\delta_{mk}X_{i}^{nl}\pm
\delta_{nl}X_{i}^{km}\right)\, .
 \label{4}
\end{equation}
The upper sign  refers to  Fermi-type operators such as  $X_{i}^{0\sigma}$, while the
lower sign refers to  Bose-type operators such as  $n_i$  (\ref{3a}) or the spin
operators (\ref{3b}). The completeness relation for the HOs, $\, X_{i}^{00} +
 \sum_{\sigma} X_{i}^{\sigma\sigma}  = 1 $,
rigorously preserves the constraint of no-double-occupancy of the quantum state
$|i,n\rangle $ on any lattice site $i$.

In terms of  HOs the  $t$-$J$  model (\ref{1}) takes  the form:
\begin{eqnarray}
H & = & - t\sum_{\langle i, j \rangle \sigma}\,  X_{i}^{\sigma 0}\, X_{j}^{0\sigma}
 - \mu \sum_{i \sigma} X_{i}^{\sigma \sigma}
\nonumber \\
&&
  +\frac{J}{4} \sum_{\langle i, j \rangle \sigma}\,
\left(X_i^{\sigma\bar{\sigma}}X_j^{\bar{\sigma}\sigma}  -
   X_i^{\sigma\sigma}X_j^{\bar{\sigma}\bar{\sigma}}\right).
\label{5}
\end{eqnarray}
We consider the $t$-$J$  model  on the honeycomb lattice  which is bipartite with two
triangular sublattices $A$ and $B$. Each of the $N$ sites on the $A$ sublattice is
connected to three nn sites $\alpha = 1, 2, 3$ belonging to the $B$ sublattice by vectors
${{\bf \delta}_\alpha}$, and $N$ sites on $B$ are connected to $A$ by vectors $-{{\bf
\delta}_\alpha}$:
\begin{equation}
{{\bf \delta}_1} = \frac{a_0}{2}(\sqrt{3}, -1),\; {{\bf \delta}_2} = -
\frac{a_0}{2}(\sqrt{3}, 1), \; {{\bf \delta}_3} = a_0(0, 1). \label{nn}
\end{equation}
The basis vectors  are ${\bf a}_1 = {{\bf \delta}_3} - {{\bf \delta}_2} =
({a_0}/{2})(\sqrt{3}, 3)$ and ${\bf a}_2 = {{\bf \delta}_3} - {{\bf \delta}_1} =
({a_0}/{2})(-\sqrt{3}, 3)$, the lattice constant is $a = |{\bf a}_1| = |{\bf a}_2| =
\sqrt{3}a_0$, where $a_0$ is  the nn distance; hereafter we put $a_0 = 1$.  The
reciprocal lattice   vectors are ${\bf k}_1 = ({2\pi}/{3})(\sqrt{3}, 1)$ and $ {\bf k}_2
=({2\pi}/{3}) (-\sqrt{3}, 1)$. In the two-sublattice representation it is convenient to
split the site indices into the unit cell and sublattice indices, $i \rightarrow  i A,
iB$.

The chemical potential $\mu$ depends on the average electron occupation number
\begin{equation}
  n = n_{A} = n_{B} = \frac{1}{N} \sum_{ i, \sigma} \langle
 \, X_{i A}^{\sigma \sigma}  \rangle ,
    \label{6}
\end{equation}
where $N$ is the number of unit cells and $\langle ...\rangle$ denotes the statistical
average with the Hamiltonian (\ref{5}).

\section{Green functions }
\label{sec:3}

\subsection{Equations for the Green functions }
\label{sec:3a}

To consider SC  within the model (\ref{5}), we introduce the anticommutator retarded
matrix GF~\cite{Zubarev60}
\begin{eqnarray}
{\sf G}_{i j \sigma} (t-t') & = &
  -i \theta(t-t') \langle \{  {\hat X}_{i \sigma }(t) ,\hat{X}^\dag_{j\sigma}(t')\}\rangle
  \nonumber \\
& \equiv & \langle \!\langle {\hat X}_{i \sigma }(t) ,{\hat X}^\dag_{j\sigma}(t')\rangle
\!\rangle,
     \label{e1}
\end{eqnarray}
where $ \{ X, Y\} = XY + YX $,  $X(t)={\rm e}^{iHt} X {\rm e}^{-iHt}$ ($\hbar =1$), and
$\,\theta(x)$ is the Heaviside function. Here we use Nambu notation and introduce the
vector Hubbard operators  $\hat{X}_{i\sigma }$ and  $\hat{X}^\dag_{j\sigma }$  with 4
components:
\begin{equation}
 \hat{X}_{i \sigma }= \left( \begin{array}{c} X_{iA}^{0 \sigma } \\
 X_{iB}^{ 0\sigma}\\  X_{iA}^{\bar{\sigma} 0} \\
 X_{iB}^{\bar{\sigma} 0} \\
  \end{array} \right),\qquad \hat{X}^\dag_{j\sigma }= \left(  X_{jA}^{\sigma 0 }\,
   X_{jB}^{\sigma 0} \, X_{jA}^{0 \bar{\sigma} }  X_{jB}^{0 \bar{\sigma} }  \right).
  \label{e2}
\end{equation}
The Fourier representation in $({\bf k}, \omega) $-space is defined by
\begin{eqnarray}
{\sf G}_{ij \sigma} (t-t')& = & \int_{-\infty}^{\infty}\frac{d\omega }{2\pi}
 e^{-i\omega(t-t')}{\sf G}_{i j \sigma} (\omega),
\nonumber \\
{\sf G}_{ij \sigma} (\omega)
 &=&\frac{1}{N}\, \sum_{\bf k}{\rm e}^{i{\bf k ({\bf r}_i-{\bf r}_j)}} {\sf G}_{\sigma}({\bf k},\omega).
     \label{e1a}
\end{eqnarray}
\par
The $4\times 4$  matrix GF (\ref{e1}) can be written as
\begin{equation}
 {\sf G}_{\sigma}({\bf k},\omega)  =
  {\hat G({\bf k},\omega)  \quad \quad
 \hat F_{\sigma}({\bf k},\omega) \choose
 \hat F^{\dagger}_{\sigma}(-{\bf k},\omega) \quad
   -\hat{G}({\bf k},-\omega) } ,
 \label{b1}
\end{equation}
where the  components  of the  normal GF  read as
\begin{eqnarray}
\hat G({\bf k},\omega)= \left(\begin{array}{cc}
 G_{AA}({\bf k}, \omega) &  G_{AB}({\bf k}, \omega) \\
   G_{BA}({\bf k}, \omega) &G_{BB}({\bf k}, \omega)
\end{array} \right) ,
\label{b2}
\end{eqnarray}
and the components  of the  anomalous  GF are given by
\begin{eqnarray}
\hat F_\sigma({\bf k},\omega)= \left(\begin{array}{cc}
 F^{\sigma}_{AA}({\bf k} \omega) &  F^{\sigma}_{AB}({\bf k}, \omega) \\
   F^{\sigma}_{BA}({\bf k}, \omega) & F^{\sigma}_{BB}({\bf k}, \omega)
\end{array} \right) .
\label{b3}
\end{eqnarray}
\par
To calculate the GF (\ref{e1}),  we use the equation of motion method. Differentiating
the GF with respect to  time $t$ we obtain
\begin{eqnarray}
 && \omega {\sf G}_{ij\sigma}(\omega) = \delta_{ij} {\sf Q} +
   \langle \langle
    [\hat X\sb{i\sigma},H] \, , \,  \hat X\sb{j\sigma}\sp{\dagger}
   \rangle \rangle_{\omega} ,
\label{e3}
\end{eqnarray}
where ${\sf Q} = \langle \{ \hat X_{i\sigma},\hat X_{i\sigma}^{\dagger}\}\rangle  =
\tilde{\tau}_{0} Q\, $. Here, $\tilde{\tau}_{0}$  is the $4 \times 4$ unit matrix and in
the paramagnetic state,  $\, Q  = \langle X\sb{i \beta}\sp{00} + X\sb{i \beta}\sp{\sigma
\sigma} \rangle = 1-n/2\, $.
\par
For a system of strongly correlated electrons as in the $t$--$J$ model it is convenient
to choose the mean-field contribution in the equations of motion (\ref{e3}) as the
zeroth-order quasiparticle (QP) energy. We calculate it in the GMFA using the  projection
operator method \cite{Plakida11}. In this approach we write the operator $ [\hat
X\sb{i\sigma}, H]$ in (\ref{e3}) as a sum of the linear part, proportional to the
single-particle  operator $\hat X_{i\sigma}$, and the irreducible part $\hat
Z\sb{i\sigma}\sp{(ir)}$ orthogonal to $\hat X_{i\sigma}$:
\begin{equation}
  \hat Z\sb{i\sigma} = [\hat X\sb{i\sigma}, H] =
    \sum\sb{l}{\sf E}\sb{il\sigma} \hat X\sb{l\sigma} +
    \hat Z\sb{i\sigma}\sp{(ir)}.
\label{e4}
\end{equation}
The orthogonality condition
 $\, \langle \{ \hat Z\sb{i\sigma}\sp{(ir)},
    \hat X\sb{j\sigma}\sp{\dagger} \} \rangle  = 0 \,$
determines  the linear part, the zeroth-order QP energy:
\begin{eqnarray}
 {\sf E}\sb{ij\sigma}=  \langle \{ [\hat X\sb{i\sigma}, H],
    \hat X\sb{j\sigma}\sp{\dagger} \} \rangle {\sf Q}^{-1} =
 { \hat{E}_{ij} \quad \quad
\hat{\Delta}_{ij\sigma} \choose \hat{\Delta}_{ji\sigma}^{*} \quad -\hat{E}_{ji} }\,  ,
\label{e5}
\end{eqnarray}
where $\hat{E}_{ij}$  and $\hat{\Delta}_{ij\sigma}$ are the normal and anomalous
components of the matrix. The corresponding zeroth-order GF in (\ref{e3})  in the Fourier
representation (\ref{e1a})  is given by
\begin{eqnarray}
  {\sf G}\sp{0}\sb{\sigma }({\bf k},\omega)& = &
    \Bigl( \omega \tilde \tau\sb{0} - {\sf E}\sb{\sigma}({\bf k})
      \Bigr) \sp{-1} {\sf Q} \, ,
  \label{e6}\\
      {\sf E}_\sigma({\bf k}) & = &
 { \hat{E}({\bf k}) \quad \quad
\hat{\Delta}_{\sigma}({\bf k}) \choose \hat{\Delta}_{\sigma}^{*}({\bf k}) \quad
-\hat{E}({\bf k}) }\,  .
 \label{e7}
\end{eqnarray}
It is possible to calculate the self-energy operator given by the many-particle GF $
\langle \langle \hat Z\sb{i\sigma}\sp{(ir)} \, \mid \,  \hat X\sb{j\sigma}\sp{\dagger}
\rangle \rangle_{\omega} $ in (\ref{e3}) and to derive the Dyson equation for the GF
(\ref{e1}), as has been done in our previous publications
(see~\cite{Plakida99,Plakida13}). In the present paper we consider the theory in GMFA
within the zeroth-order approximation for the GF (\ref{e6}).

The components of the energy matrix (\ref{e5})  are determined by the commutators:
\begin{eqnarray}
  \hat{E}_{ij}& = & \langle \{ [ \left( \begin{array}{cc} X_{iA}^{0 \sigma } \\
 X_{iB}^{ 0\sigma} \\
  \end{array} \right), H], \left(  X_{jA}^{\sigma 0 }\,
   X_{jB}^{\sigma 0}   \right)
    \} \rangle {\sf Q}^{-1}  ,
    \label{a1}\\
   \hat{\Delta}_{ij,\sigma}& = & \langle \{ [ \left( \begin{array}{cc} X_{iA}^{0 \sigma } \\
 X_{iB}^{ 0\sigma} \\
  \end{array} \right), H], \left(   X_{jA}^{0 \bar{\sigma} }  X_{jB}^{0 \bar{\sigma} }   \right)
    \} \rangle {\sf Q}^{-1}.
 \label{a2}
\end{eqnarray}
Performing  commutations and introducing the Fourier representation, $X_{iA}^{0 \sigma}=
({1}/{ \sqrt{N}}) \sum_{\bf k}{\rm e}^{i{\bf k} {\bf r}_i} X_{{\bf k} A}^{0 \sigma}\,$,
we obtain:
\begin{eqnarray}
  \hat{E}({\bf k})&=&
  \left(
\begin{array}{cc}
\varepsilon_A  &  \varepsilon_{AB}({\bf k}) \\
     \varepsilon_{BA}({\bf k}) & \varepsilon_B
\end{array} \right) ,
 \label{a3}\\
 \varepsilon_A  &= &  \langle \{ [ X_{{\bf k} A}^{0 \sigma }, H],
X_{{\bf k} A}^{\sigma 0}\} \rangle \; {Q}^{-1}=  -\widetilde{ \mu } ,
\nonumber\\
 \varepsilon_B &= & \langle \{ [ X_{{\bf k} B}^{0 \sigma }, H],
X_{{\bf k} B}^{\sigma 0}\} \rangle \; {Q}^{-1} = -\widetilde{ \mu },
 \nonumber\\
 \varepsilon_{AB}({\bf k})  &= &  \langle \{ [ X_{{\bf k} A}^{0 \sigma }, H],
X_{{\bf k} B}^{\sigma 0}\} \rangle \; {Q}^{-1} = - \widetilde{t}\, \gamma ({\bf k}),
\nonumber\\
  \varepsilon_{BA}({\bf k})& = &  \varepsilon_{AB}({\bf k})^*    = - \widetilde{t}\, \gamma^*({\bf k}),
 \label{a4}
\end{eqnarray}
where  $\gamma({\bf k}) = \sum_\alpha \exp (i {\bf k} \overrightarrow{\delta_\alpha})$
and $ |\gamma({\bf k})|^2 = {1 + 4 \cos ( \sqrt{3}k_x /2)[\cos (\sqrt{3} k_x /2) + \cos
({3}k_y/{2})]} $. The renormalized chemical potential $\widetilde{ \mu } $ and  hopping
parameter $ \widetilde{t} $ were calculated in Ref.~\cite{Vladimirov18} and are given by
the relations:
\begin{eqnarray}
&& \widetilde{ \mu }  =
   \mu -  \frac{3t}{Q}\, D_1 + \frac {3J}{4} n - \frac {3J}{2Q}\, C_1 ,
 \label{a5}\\
&& \widetilde{t}  =  t\, Q \, \left( 1 + \frac{3 C_1}{2Q^2}\right) + J\frac { D_1}{2 Q}.
\label{a6}
\end{eqnarray}
Here the nn correlation functions   for electrons and spins are:
\begin{eqnarray}
D_1 = \langle X^{\sigma 0}_{iA} X^{0\sigma}_{i+\delta_1, B}\rangle, \quad C_1 = \langle
S^z_{iA} {S}^z_{i+\delta_1, B} \rangle .
 \label{a6a}
\end{eqnarray}

The anomalous energy matrix for the gaps  reads:
\begin{eqnarray}
 &&\hat{\Delta}_{\sigma}({\bf k})=
   \left(
\begin{array}{cc}
\Delta_{A \sigma}  &  \Delta_{AB\sigma}({\bf k}) \\
    \Delta_{BA\sigma}({\bf k}) &\Delta_{ B\sigma}
\end{array} \right) ,
\label{a7}\\
&& \Delta_{A\sigma}= \Delta_{B\sigma} \equiv \Delta_{\sigma}=
 \langle \{ [ X_{{\bf k} A}^{0 \sigma }, H], X_{-{\bf k} A}^{0
\bar\sigma }\} \rangle \; {Q}^{-1}
\nonumber \\
&&  \quad = 2 t\sum_{ l}\, \langle  X_{lB}^{0 \bar{\sigma} }X_{iA}^{0 \sigma}\rangle
{Q}^{-1} ,
  \label{a8}\\
&& \Delta_{AB\sigma}({\bf k})  =  \langle \{ [ X_{{\bf k} A}^{0 \sigma }, H], X_{-{\bf k}
B}^{0 \bar\sigma } \} \rangle \; {Q}^{-1}
\nonumber \\
   & = & - \frac{J}{Q}\sum_{{\bf r_j= r_i + \delta_\alpha}} \exp[i{\bf k (r_j-r_i)}]\,
 \langle X_{j B}^{0 \bar{\sigma}}   X_{i A}^{0 {\sigma}}\rangle ,
  \label{a9}\\
&& \Delta_{BA\sigma}({\bf k}) =\langle \{ [ X_{{\bf k} B}^{0 \sigma }, H], X_{-{\bf k}
A}^{0 \bar\sigma } \} \rangle \; {Q}^{-1}
 \nonumber \\
   & = &  -  \Delta_{AB \bar\sigma }(- {\bf k}) =
 \Delta_{AB \sigma }(- {\bf k}).
  \label{a10}
 \end{eqnarray}
Note that both gap functions $\Delta_{\sigma}$ and  $\Delta_{AB\sigma}({\bf k})$ are
determined by the nn correlation functions $\langle X_{j B}^{0 \bar{\sigma}}   X_{i A}^{0
{\sigma}}\rangle$, since we have no pairing on one lattice site contrary to
Ref.~\cite{Uchoa07}.
\par
Using Eqs.~(\ref{a3})  and (\ref{a7}) we obtain the energy matrix:
\begin{eqnarray}
{\sf E}_\sigma({\bf k})= \left( \begin{array}{cccc}
 -\widetilde{ \mu } \quad - \widetilde{t}\, \gamma ({\bf k})\quad
\Delta_{\sigma}  \quad  \Delta_{AB\sigma}({\bf k})
 \\
- \widetilde{t}\, \gamma^*({\bf k}) \quad -\widetilde{ \mu }\quad
 \Delta_{AB\sigma}(-{\bf k})\quad \Delta_{\sigma}\quad
\\
\Delta^*_{\sigma}  \quad  \Delta^*_{AB \sigma}(-{\bf k})\quad \widetilde{ \mu } \quad
\widetilde{t}\, \gamma ({\bf k})\quad
\\
 \Delta^*_{AB \sigma}({\bf k}) \quad \Delta^*_{\sigma} \quad
\widetilde{t}\, \gamma ^*({\bf k}) \quad \widetilde{ \mu }
\\
  \end{array} \right). \quad
       \label{e8}
\end{eqnarray}
We point out that the matrix (\ref{e8}) is similar to the matrix in the superconducting
state in MFA for graphene obtained in Ref.~\cite{Uchoa07}  and in Ref.~\cite{Jiang08} but
with the renormalized chemical potential $\widetilde{ \mu }$   and hopping parameter
$\widetilde{t}$.

The GF in  Eq.~(\ref{e6})  is defined by the inverse matrix $\, \Bigl( \omega \tilde
\tau\sb{0} - {\sf E}\sb{\sigma}({\bf k})   \Bigr) \sp{-1}$. Its  calculation results in
the GF:
\begin{eqnarray}
&& {\sf G}\sp{0}\sb{\sigma }({\bf k},\omega)
 = \frac{Q}{\widetilde{D}({\bf k},\omega)} {\sf A}^{\rm T}_{\sigma }({\bf k},\omega),
\label{e10}
\end{eqnarray}
where ${\sf A}^{\rm T}_{\sigma }({\bf k},\omega)$ is   the transposed matrix of the
cofactors of the matrix $ \Bigl( \omega \tilde \tau\sb{0} - {\sf E}\sb{\sigma}({\bf k})
\Bigr)$.

The diagonal and off-diagonal normal $ G_{\alpha\beta}({\bf k}, \omega)$ and anomalous  $
F_{\alpha\beta\sigma}({\bf k}, \omega) $  GFs components in the matrix  (\ref{b1}) read:
\begin{eqnarray}
 G_{AA}({\bf k}, \omega)& = &  \langle \langle
    X^{0 \sigma}_{{\bf k} A}\, ,\,  X^{ \sigma 0}_{{\bf k} A}   \rangle \rangle_{\omega}
     = \frac{ A_{11}Q}{\widetilde{D}({\bf k},\omega)} ,
 \label{e12a1} \\
G_{AB}({\bf k}, \omega) & = & \langle \langle
    X^{0 \sigma}_{{\bf k} A}\, ,\,  X^{ \sigma 0}_{{\bf k} B}   \rangle \rangle_{\omega}
    = \frac{ A_{21}Q}{\widetilde{D}({\bf k},\omega)},
 \label{e12a}\\
 F_{AA}^{\sigma}({\bf k}, \omega) & = & \langle \langle
    X^{0 \sigma}_{{\bf k} A}\, ,\,  X^{ 0 \bar{\sigma}}_{-{\bf k} A}   \rangle \rangle_{\omega} =
    \frac{ A_{31}^{\sigma} Q}{\widetilde{D}({\bf k},\omega)}, \;
 \label{e12b1} \\
 F_{AB}^{\sigma}({\bf k}, \omega) & = & \langle \langle
    X^{0 \sigma}_{{\bf k} A}\, ,\,  X^{ 0 \bar{\sigma}}_{-{\bf k} B}   \rangle \rangle_{\omega} =
     \frac{ A_{41}^{\sigma} Q}{\widetilde{D}({\bf k},\omega)} .
 \label{e12b}
\end{eqnarray}
The coefficients $A_{nm}$ are given by the equations:
\begin{eqnarray}
 && A_{11}({\bf k})= (\omega^2 - \widetilde{ \mu }^2)(\omega - \widetilde{ \mu})
 - \Delta^*_{AB \sigma}(-{\bf k})\, \Delta^*_{\sigma}\,\widetilde{t} \gamma ({\bf k})
\nonumber\\
 &&   - \Delta_{\sigma} \Delta^*_{AB \sigma}(-{\bf k})\,
\widetilde{t}\, \gamma ^*({\bf k})- |\Delta_{\sigma})|^2\,( \omega -\widetilde{ \mu})
\nonumber\\
 &&-(\omega +\widetilde{ \mu })\,
\widetilde{t}^2\, |\gamma ({\bf k})|^2
 -  | \Delta_{AB\sigma}(-{\bf k})|^2 (\omega - \widetilde{ \mu}),
\label{b4}\\
&& A_{21}({\bf k}) = - \widetilde{t}\, \gamma ({\bf k})\, (\omega -\widetilde{ \mu })^2 +
|\Delta_{\sigma}|^2 \, \widetilde{t}\, \gamma ({\bf k})
\nonumber\\
 && + \Delta_{AB\sigma}^2({\bf k}) \, \widetilde{t}\, \gamma ^*({\bf k}) + \Delta_{AB\sigma}({\bf k})\,\Delta^*_{ \sigma} \, (\omega -\widetilde{ \mu })
\nonumber\\
&& + \widetilde{t}^3\, \gamma ({\bf k})\,|\gamma ({\bf k}))|^2
 + \Delta_{\sigma}\, \Delta_{AB\sigma}({\bf k})\,  (\omega - \widetilde{ \mu }) ,
 \label{b5}\\
&&  A_{31}^{\sigma}({\bf k})  = - \Delta_{AB\sigma}(-{\bf k})\, \widetilde{t}\, \gamma
({\bf k})\,( \omega - \widetilde{ \mu })
 - \Delta_{\sigma}\,| \Delta_{B\sigma} |^2
 \nonumber\\
&&  +  \Delta_{AB\sigma}({\bf k}) \, \widetilde{t}\, \gamma ^*({\bf k})\,(\omega
+\widetilde{ \mu })
 + \Delta_{AB\sigma}({\bf k}) \, \Delta_{AB\sigma}(-{\bf k})\,\Delta^*_{\sigma}
 \nonumber\\
 &&    + \Delta_{\sigma}\, ( \omega^2  - \widetilde{ \mu }^2 - \widetilde{t}^2\, |\gamma ({\bf k})|^2) ,
 \label{b6}\\
&&  A_{41}^{\sigma}({\bf k}) =  - \Delta_{AB\sigma}(-{\bf k})\, \widetilde{t}^2\,
\gamma^2 ({\bf k})
  + \Delta^2_{\sigma}\,   \Delta_{AB\sigma}({\bf k})
 \nonumber\\
&& + \Delta_{AB\sigma}({\bf k}) (\omega^2 -\widetilde{ \mu }^2 ) -
\Delta_{AB\sigma}^2({\bf k})\,  \Delta_{AB\sigma}(-{\bf k})
 \nonumber\\
 &&    + \Delta_{\sigma}  \, 2\widetilde{ \mu } \, \widetilde{t}\, \gamma ({\bf k}) .
   \label{b7}
\end{eqnarray}

 The determinant $\widetilde{D}({\bf k},\omega)$  of the matrix,
\begin{eqnarray}
 \widetilde{D}({\bf k},\omega) = \parallel \omega \tilde \tau\sb{0} - {\sf E}\sb{\sigma}({\bf k}) \parallel,
 \label{e10a}
\end{eqnarray}
gives the equation for the spectrum in the superconducting state $\widetilde{D}({\bf
k},\omega) = 0$ which can be written in a general case (with notations $\Delta_{\sigma
\pm} \equiv |\Delta_{AB \sigma}({\pm \bf k})|$) as:
\begin{eqnarray}
&&\omega^4 - 2\omega^2(\Delta_{\sigma}^2 + \frac{1}{2}(\Delta_{\sigma +}^2 +
\Delta_{\sigma -}^2) +  \widetilde{t}^2\, |\gamma ({\bf k})|^2 + \widetilde{ \mu }^2)
\nonumber\\
&& + \Delta_{\sigma +}^2 \, \Delta_{\sigma -}^2
 + \widetilde{t}^4 \, |\gamma ({\bf k})|^4
 + \widetilde{ \mu }^4 + \Delta_{\sigma}^4 \nonumber\\
&& + 2\, \widetilde{t}^2 \, {\rm Re}[\gamma ({\bf k})^2 \Delta^*_{AB \sigma}({\bf
k})\Delta_{AB \sigma}(-{\bf k})] - 2\widetilde{t}^2\, |\gamma ({\bf k})|^2 \widetilde{
\mu }^2
\nonumber\\
&&+ (\Delta_{\sigma +}^2 + \Delta_{\sigma -}^2) \, \widetilde{ \mu}^2
  + 2\Delta_{\sigma}^2 (\widetilde{ \mu}^2 +  \widetilde{t}^2 \, |\gamma ({\bf k})|^2)
   \nonumber\\
&&- 4 \, \widetilde{t}\, \widetilde{ \mu } \, \Delta_{\sigma} \, {\rm Re}[\Delta_{AB
\sigma}(-{\bf k})  \, \gamma ({\bf k}) + \Delta_{AB\sigma}({\bf k}) \, \gamma^*({\bf k})]
\nonumber\\
&& - 2\, \Delta_{\sigma}^2 {\rm Re} [\Delta_{AB\sigma}({\bf k}) \Delta_{AB\sigma}(-{\bf
k})]= 0 .
 \label{b8}
\end{eqnarray}
The solution of this equation  for the extended $s$-pairing (see Eq.~(\ref{e17})) gives
the spectrum of excitations:
\begin{eqnarray}
&&\Omega^2({\bf k})  =  (\widetilde{ \mu } \pm \widetilde{t}\, |\gamma ({\bf k})|)^2 +
(\Delta_\sigma \pm |\Delta_{AB\sigma}({\bf k})|)^2, \qquad
 \label{e11}
\end{eqnarray}
which coincides with the spectrum for graphene in MFA found in Ref.~\cite{Uchoa07}. Note
that for the gap $\Delta_\sigma  = 0$ the spectrum is gapless at $\widetilde{ \mu }= 0$
at six corners  of the BZ at $K, K'$ points:  $\Omega_s({\bf k}) = \widetilde{t}\,|\gamma
({\bf k})|\sqrt{ 1 + \Delta^2_{AB\sigma}/\widetilde{t}^2 } $, as was pointed out
in~\cite{Uchoa07}.

The spectrum for the $d$-pairing (see Eq.~(\ref{e18}))  resulting from Eq.~(\ref{b8})
with $\Delta_\sigma = 0 $ has a more complicated structure with  two gaps $\Delta_{\sigma
\pm} $:
\begin{eqnarray}
&&\Omega^2({\bf k}) = \frac{1}{2}(\Delta_{\sigma +}^2 + \Delta_{\sigma -}^2) +
\widetilde{t}^2\, |\gamma ({\bf k})|^2 + \widetilde{ \mu }^2
 \nonumber\\
&&\pm \Big\{\widetilde{t}^2\, |\gamma ({\bf k})|^2 (4\,\widetilde{ \mu }^2 +
\Delta_{\sigma +}^2 + \Delta_{\sigma -}^2)+ \frac{1}{4}(\Delta_{\sigma +}^2 -
\Delta_{\sigma -}^2)^2
  \nonumber\\
 & & - 2\, \widetilde{t}^2 \, {\rm Re}[\gamma ({\bf k})^2 \Delta^*_{AB \sigma}({\bf k})\Delta_{AB \sigma}(-{\bf k})]
\Big\}^{1/2}.
  \label{e11a}
\end{eqnarray}
A similar spectrum was found for the $d$-wave  symmetry in Ref.~\cite{Jiang08}. Since the
$d$-wave gap (\ref{e18}) is zero at three corners of the BZ at  $K$  points and has a
maximum  value at another three corners of the BZ at $K'$ points,  the spectrum
(\ref{e11a}) is gapless at $\widetilde{ \mu }= 0$  at  $K$ points and has a maximum value
at  $K'$ points (see also Refs.~\cite{Kotov12,Wu13}).

\subsection{Normal state Green function }
\label{sec:3b}

The determinant for the normal state GFs  (\ref{e12a1}) and (\ref{e12a}) reads
\begin{eqnarray}
{\widetilde{D}({\bf k},\omega)} & = & D({\bf k}, \omega)\,  D({\bf k}, -\omega),
 \nonumber \\
D({\bf k}, \omega)  &= & [\varepsilon_{+}({\bf k}) - \omega] [\varepsilon_{-}({\bf k}) -
\omega].
 \label{e14}
\end{eqnarray}
Here the electronic spectrum  is given by the  matrix (\ref{e8}) in the normal state (cf.
Ref.~\cite{Vladimirov18}):
\begin{eqnarray}
\varepsilon_{\pm}({\bf k}) = -\widetilde{ \mu } \pm  \widetilde{t}\,|\gamma({\bf k})| .
\label{e9}
\end{eqnarray}
The spectrum has the Dirac cone-type behaviour at  $K$ and $K'$ points at the corners of
the BZ as in graphene. The cones  touch the Fermi surface (FS) at $\widetilde{ \mu } = 0$
for half filling at  the electron occupation number $n = 2/3$ in the $t$-$J$  model. The
detailed doping dependence of the FS was  considered  in Ref.~\cite{Vladimirov18}.

For the normal state GFs  (\ref{e12a1}) and  (\ref{e12a}) with the coefficients $A_{11}$
(\ref{b4}) and $A_{21}$ (\ref{b5})   we obtain:
  \begin{eqnarray}
 G_{AA}({\bf k}, \omega)
=  \frac{Q}{2} \Big[ \frac{1}{\omega - \varepsilon_+({\bf k})} +
 \frac{1}{\omega - \varepsilon_-({\bf k})} \Big] ,
\label{e13a}
\end{eqnarray}
\begin{eqnarray}
 G_{AB}({\bf k}, \omega) &= &  - \frac{ Q \gamma ({\bf k})\,}{ 2 \,| \gamma ({\bf k})| }
 \Big[ \frac{1}{\omega - \varepsilon_+({\bf k})} -
  \frac{1}{\omega - \varepsilon_-({\bf k})} \Big]. \quad
\label{e13b}
\end{eqnarray}
The density of electronic states (DOS) at the Fermi energy $\widetilde{\mu}$ in the
normal state is determined by the GF (\ref{e13a}), and for the dispersion
$\varepsilon_{\pm}({\bf k})$ (\ref{e9}), is given by the equation:
  \begin{eqnarray}
N(\widetilde{\mu})& = &\frac{1}{N} \sum_{{\bf k},\sigma}[-\frac{1}{\pi}]\, {\rm Im}\, G_{AA}({\bf k},
0+ i\epsilon)
  \nonumber \\
 & = & \frac{Q}{N} \sum_{{\bf k}}[ \delta( \widetilde{ \mu } - \widetilde{t}\,|\gamma({\bf k})|) + \delta(  \widetilde{ \mu } + \widetilde{t}\,|\gamma({\bf k})|)].
  \label{13c}
\end{eqnarray}
\begin{figure}
\centering
\resizebox{0.4\textwidth}{!}{%
\includegraphics{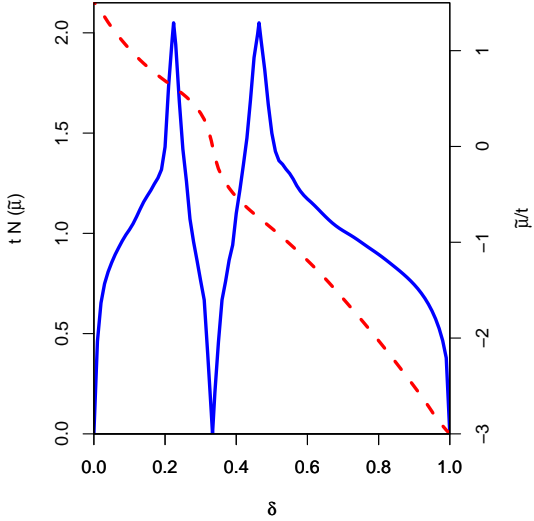}}
\caption{(Color online) Density of electronic states at the Fermi energy $t\,
N(\widetilde{ \mu })$ (solid blue line, left axis) and the chemical potential $
\widetilde{ \mu }/t $ (dashed red line, right axis)  {\it vs} doping $\delta$.  }
 \label{fig1}
\end{figure}
The DOS at the Fermi energy  $N(\widetilde{\mu})$  and the  chemical potential $\widetilde{ \mu }$ for
$ -3 \leq \widetilde{\mu}/t \leq 1.5$ at zero temperature are plotted in Fig.~\ref{fig1}
for doping $0 \leq \delta \leq 1$. The two peaks of the DOS correspond to the VHS in two
bands. The DOS is nonsymmetric with respect to $\widetilde{ \mu }= 0$ in comparison with
the DOS of  graphene due to the renormalization of the hopping parameter in (\ref{13c}),
where we use $\widetilde{t} = t \,Q = t\, (1+\delta)/2$ neglecting the electron and spin
contributions in Eq.~(\ref{a6}).

We note that there are  misprints in Ref.~\cite{Vladimirov18},  Eqs.~(26) and  (29) for
the  GFs, in comparison with Eqs.~(\ref{e13a}) and (\ref{e13b}),  where we have opposite
signs. But  the expressions for the correlation functions $  n_{A \sigma}({\bf k}) =
\langle X_{{\bf k}A }^{\sigma 0}  X_{{\bf k}A}^{0 \sigma} \rangle$ and $  \langle X_{{\bf
k} B }^{\sigma 0}  X_{{\bf k} A}^{0 \sigma} \rangle $  in Ref.~\cite{Vladimirov18} are
correct.

\subsection{Anomalous Green functions}
\label{sec:3c}

To calculate  the superconducting $T_c$ we  use the linearized approximation for the
anomalous GFs (\ref{e12b1}) and (\ref{e12b}). They are determined by the relations:
\begin{eqnarray}
&& F_{AA}^{\sigma}({\bf k}, \omega) =
   \frac{Q \, A_{31}^{ \sigma}({\bf k}, \omega)}{ [\omega^2 -\varepsilon^2_{+}({\bf k}) ]
 [\omega^2 -\varepsilon^2_{-}({\bf k})  ]},
\label{e15}\\
&&  A_{31}^{ \sigma}({\bf k}, \omega)
 = - \Delta_{AB\sigma}(-{\bf k})\,( \omega - \widetilde{ \mu })\, \widetilde{t}\, \gamma ({\bf k})
 \nonumber\\
 && +  \Delta_{AB\sigma}({\bf k}) (\omega +\widetilde{ \mu }) \,
 \widetilde{t}\, \gamma ^*({\bf k})
 \nonumber\\
 &&    + \Delta_{\sigma} \,( \omega^2  - \widetilde{ \mu }^2  - \widetilde{t}^2\, |\gamma ({\bf k})|^2),
 \label{e15a}
\end{eqnarray}
\begin{eqnarray}
&& F_{AB}^{\sigma}({\bf k}, \omega)=
   \frac{Q \, A_{41}^{ \sigma}({\bf k}, \omega)}{ [\omega^2 -\varepsilon^2_{+}({\bf k}) ]
 [\omega^2 -\varepsilon^2_{-}({\bf k})  ]},
\label{e16}\\
&& A_{41}^{ \sigma}({\bf k}, \omega) = \Delta_{AB\sigma}({\bf k}) (\omega^2 -\widetilde{
\mu }^2 )
  \nonumber\\
 &&- \Delta_{AB\sigma}(-{\bf k}) \, \widetilde{t}^2\, \gamma^2 ({\bf k})
+ \Delta_{\sigma} \, 2 \widetilde{ \mu } \, \widetilde{t}  \, \gamma ({\bf k}) .
\label{e16a}
\end{eqnarray}
The corresponding   anomalous correlation function  $\, F_{B A}^{{\sigma}}({\bf k}) =
\langle X^{ 0 \bar{\sigma}}_{-{\bf k} B}
    X^{0 \sigma}_{{\bf k} A} \rangle  \,$ in  the gap equations (\ref{a8}),
(\ref{a9}) determined by the GF (\ref{e16})  is given by
 \begin{eqnarray}
 && F_{B A}^{{\sigma}}({\bf k})
 =  - \frac{1}{\pi}\int_{-\infty}^{\infty} \frac{d\omega }{\exp(\omega/T)+ 1}\;
 {\rm Im} F_{AB}^{\sigma}({\bf k}, \omega)
   \nonumber\\
    &= & \frac{Q \,}{ \varepsilon^2_{-}({\bf k})-
\varepsilon^2_{+}({\bf k})}
 \Bigl\{
\frac{ A^{\sigma}_{41}(\omega = \varepsilon_{+}({\bf k}) )}{2 \varepsilon_{+}({\bf k})}
 \tanh \frac{\varepsilon_{+}({\bf k})}{2T}
  \nonumber\\
 &- &  \frac{ A^{\sigma}_{41}(\omega = \varepsilon_{-}({\bf k}) )}{2 \varepsilon_{-}({\bf k})}
  \tanh \frac{\varepsilon_{-}({\bf k})}{2T}  \Bigr\}.
    \label{e19}
\end{eqnarray}

\section{Gap equations and $\bf T_c $}
\label{sec:4}

Let us consider  solutions for the gaps of different symmetries. In particular,  the
extended $s$-wave gap (\ref{a9}) determined by the   bond-independent anomalous
correlation function (\ref{e19}) can be written as:
   \begin{eqnarray}
\Delta_{AB\sigma}({\bf k}) =\Delta_{AB\sigma}\, \sum_{{\alpha}} \exp[i{\bf (k
\delta_{\alpha}}] = \Delta_{AB\sigma}\,  \gamma ({\bf k}) .
 \label{e17}
 \end{eqnarray}
The $ d+id'$-wave pairing is determined by the gap which has different phases on bonds
(Refs.~\cite{Black07,Jiang08,Gu13}):
   \begin{eqnarray}
\Delta_{AB\sigma}({\bf k}) = \sum_{\alpha}\Delta^{\alpha}_{\sigma} \exp[i{\bf k
\delta_{\alpha}}],\;  \Delta^{\alpha}_{\sigma} = \Delta_{1 \sigma}e^{i(2\pi/3)\alpha}.
 \label{e18}
 \end{eqnarray}
The numerical solution of the gap equations   yields $T_c$ as a function of doping given
below, where  we use the renormalized hopping parameter $
\widetilde{t} = t\,(1+\delta)/2$  (see Sect.~\ref{sec:3b}).  Note that in our mean-field approach in two dimensions, $T_c$ is the
temperature of Cooper-pair formation without superconducting long-range order (see, e.g.,
Refs.~\cite{Kotov12,Uchoa07}).

It should be pointed out  that for the conventional Fermi-liquid we can have the $s$-wave
pairing on a single site given  by the anomalous correlation function $\langle a_{i
\bar\sigma}  a_{i \sigma} \rangle$. In terms of  HOs this pairing can be described by the
equation:
\begin{eqnarray}
\langle a_{i \bar\sigma}  a_{i \sigma} \rangle = \langle X_{i} ^{02}  \rangle =\langle
X_{i} ^{0\bar\sigma} X_{i} ^{\bar\sigma 2} \rangle \neq 0,
     \label{c1a}
\end{eqnarray}
which shows that the double occupancy of one lattice site is permitted but in the  two
Hubbard subbands. For the Hubbard model in the limit of strong correlations, i.e.,  for
the $t$-$J$ model,   this pairing is prohibited due to  the no-double-occupancy
restriction which in terms of  HOs,  as was proposed in Ref.~\cite{Plakida89}, can be
written as:
\begin{eqnarray}
\langle    X_{i }^{0 \bar{\sigma} }X_{i}^{0 \sigma}\rangle = \langle
a_{i\bar{\sigma}}(1-n_{i ,{\sigma}})\, a_{i\sigma}(1-n_{i ,\bar{\sigma}})\rangle
 \equiv 0.
\label{c1}
\end{eqnarray}
As shown in  Sect.~\ref{sec:c}, both the ${\bf k}$-independent $s$-wave pairing with the
gap (\ref{a8}) and the extended $s$-wave pairing with the gap (\ref{e17})  violate this
condition and must be excluded from a rigorous point of view.  For the  $d$-pairing with
the gap (\ref{e18}) the condition (\ref{c1}) is fulfilled.

However, in several publications this constraint has not been rigorously taken into
account, for example using the phenomenological $t$-$J$ model (see  Ref.~\cite{Black07}
where the  double-site occupancy is  included in the Hilbert space), or introducing the
statistical weighting factors $g_t = 2\delta/(1 + \delta)$ and $g_J = 4/(1 + \delta)^2$
in the $t$-$J$ model  (see, Refs.~\cite{Black14,Wu13}), as discussed in
Sect.~\ref{sec:1}, or using the Gutzwiller projector $ P_g =\prod_i(1- g\, n_{i \uparrow}
n_{i \downarrow} )$ with   $ \, g < 1\,$ treated as variational parameter (see, e.g.,
Ref.~\cite{Pathak10}). To compare our results with these studies   we have considered the
${\bf k}$-independent $s$-wave pairing  in Sect.~\ref{sec:4a} and the extended $s$-wave
pairing  in Sect.~\ref{sec:4b}.

\subsection{ ${\bf k}$-independent $s$-wave gap equation}
\label{sec:4a}

For the ${\bf k}$-independent  $s$-wave  gap  (\ref{a8}) we have
\begin{eqnarray}
 \Delta_{\sigma} &= &
    \frac{2t}{Q}\sum_{ l}\,\langle X_{lB}^{0\bar{\sigma}} X_{i A}^{ 0 \sigma}\rangle
 \nonumber \\
 & = &\frac{2t}{Q N} \sum_{{\alpha}} \sum_{\bf q} \exp[-i{\bf  q \delta_{\alpha}}] \,
 F_{B A}^{\sigma}({\bf q}).
 \label{e20}
\end{eqnarray}
 Assuming that $t \gg J$ we can solve the equation for $\Delta_{ \sigma}$
 considering the ${\bf q}$-dependent  part of the gap
$\Delta_{AB \sigma}({\bf q})$ in $ A^{\sigma}_{41}(\omega,{\bf q} )$ (\ref{e16a})
as a small perturbation. We obtain the gap equation:
\begin{eqnarray}
\Delta_{\sigma}  &= & \frac{\Delta_{\sigma} \, t}{N} \,\sum_{\bf q} |\gamma({\bf q})|
 \Bigl[\frac{ 1}{2 \varepsilon_{+}({\bf q})}
 \tanh \frac{\varepsilon_{+}({\bf q})}{2T_c}
\nonumber\\
 &- & \frac{ 1}{2 \varepsilon_{-}({\bf q})}
 \tanh \frac{\varepsilon_{-}({\bf q})}{2T_c}\Bigr] \Bigr\}.
 \label{e21}
\end{eqnarray}
We find a solution for $T_c $  only for  hole doping $\delta \leq 0.32$ when $\tilde{\mu}
> 0$.  It has a high maximum value, $T_c \sim 0.25 t$, due to the strong pairing
interaction $t \gg J$. Note that the coupling proportional to the hopping parameter $t$
is due to the kinematical interaction induced by  the commutation relations (\ref{4}) for
the  HOs.  The same type of pairing occurs in the $t$--$J$ model on the square lattice
(see Ref.~\cite{Plakida89}) which has been disregarded, since it violates the constraint
of  no-double-occupancy (\ref{c1}).  As shown in Sect.~\ref{sec:c}, the $s$-wave pairing
described by the ${\bf k}$-independent gap $\Delta_{ \sigma}$ (\ref{e20}) on the
honeycomb lattice also violates the constraint (\ref{c1}),  and in what follows we take
the solution  $ \Delta_{ \sigma} =0 $.

\subsection{Extended $s$-wave gap equation}
\label{sec:4b}

Using Eq.~(\ref{e19}) the gap equation   (\ref{a9}) reads
 \begin{eqnarray}
&& \Delta_{AB\sigma}({\bf k}) =   \frac{J}{ 2 N} \sum_{\bf q} \sum_{{\alpha}} \exp[i{\bf
(k- q ) \delta_{\alpha}}]
   \Delta_{AB\sigma}({\bf q})
 \nonumber\\
 &\times &
 \Bigl[\frac{ 1}{2 \varepsilon_{+}({\bf q})}
 \tanh \frac{\varepsilon_{+}({\bf q})}{2T_c} + \frac{ 1}{2 \varepsilon_{-}({\bf q})}
 \tanh \frac{\varepsilon_{-}({\bf q})}{2T_c}\Bigr]
  \nonumber\\
 & - &\frac{J}{ N} \sum_{\bf q} \sum_{\alpha} \exp[i{\bf (k- q ) \delta_{\alpha}}]\frac{ \widetilde{t}^2}{4 \tilde{\mu}\, \tilde{t}|\gamma({\bf q})| }
 \nonumber\\
 & & \times \Bigl[
\Delta_{AB\sigma}({\bf q})  |\gamma({\bf q})|^2 -
   \Delta_{AB\sigma}(-{\bf q}) \, \ \gamma^2 ({\bf q}) \Bigr]
 \nonumber\\
& \times & \Bigl[\frac{ 1}{2 \varepsilon_{+}({\bf q})}
 \tanh \frac{\varepsilon_{+}({\bf q})}{2T_c} - \frac{ 1}{2 \varepsilon_{-}({\bf q})}
 \tanh \frac{\varepsilon_{-}({\bf q})}{2T_c}\Bigr].
    \label{e22}
\end{eqnarray}
Taking into account that the second term  vanishes for the bond-independent gap
$\Delta_{AB\sigma}({\bf k})
 = \Delta_{AB\sigma}\,\gamma({\bf k})$ because of   $\, \gamma({\bf q})  |\gamma({\bf q})|^2 -
   \gamma^*({\bf q}) \, \ \gamma^2 ({\bf q}) = 0 \,$, we obtain the  gap equation
\begin{eqnarray}
&&   \gamma ({\bf k})
   =   \frac{J}{2 N} \sum_{\bf q} \sum_{\alpha}  \exp[i{\bf (k- q ) \delta_{\alpha}}]
     \gamma ({\bf q})
 \nonumber\\
 &\times &
 \Bigl[\frac{ 1}{2 \varepsilon_{+}({\bf q})}
 \tanh \frac{\varepsilon_{+}({\bf q})}{2T_c} + \frac{ 1}{2 \varepsilon_{-}({\bf q})}
 \tanh \frac{\varepsilon_{-}({\bf q})}{2T_c}\Bigr] .
    \label{e23}
\end{eqnarray}
The solution of this equation shows  that $T_c(\delta)$ linearly increases with doping
and vanishes for large $\,\delta\,$ depending on $J$, e.g., $T_c(\delta) = 0$ at  $
\delta > 0.2$  for $J = t/2$. The maximum value   of $T_c$ rapidly  increases with the
interaction $J$, e.g., $T_c^{\rm max} = 0.035\, t \; (0.082\, t )\,$ for   $J =  t /3\;
(\, t/2 )$. Though the extended $s$-wave pairing  also  violates the constraint of
no-double-occupancy (\ref{c1}), we consider it to compare our results with calculations
for the $t$-$J$ model, where this restriction has  not been rigorously taken into
account.

\subsection{$d$-wave gap equation}
\label{sec:4c}

In the case of   $d$-wave symmetry  ($d +i d'$) for the bond-dependent   gap (\ref{e18}) we
have the same  equation (\ref{e22}).  The direct numerical calculation  for this gap
reveals that the second term in the integral gives no contribution. It is convenient to
consider the  equation for a particular value of $\alpha$ for the phase sensitive
contribution:
\begin{eqnarray}
&& \Delta^{\alpha}_{\sigma} \exp[i{\bf (k \delta_{\alpha}}]  =  \Delta_{1\sigma}
e^{i(2\pi/3)\alpha} \exp[i{\bf k  \delta_{\alpha}}]
 \nonumber\\
&&   =   \frac{J}{ 2N} \sum_{\bf q} \exp[i{\bf (k - q) \delta_{\alpha}}]
    \sum_{\beta}\,  \Delta_{1\sigma} \, e^{i(2\pi/3)\beta} \exp[i{\bf q  \delta_{\beta}}]
 \nonumber\\
 &\times &
 \Bigl[\frac{ 1}{2 \varepsilon_{+}({\bf q})}
 \tanh \frac{\varepsilon_{+}({\bf q})}{2T_c} + \frac{ 1}{2 \varepsilon_{-}({\bf q})}
 \tanh \frac{\varepsilon_{-}({\bf q})}{2T_c}\Bigr] .
    \label{g32a}
\end{eqnarray}
Cancelling out the  terms $\Delta_{1\sigma}\,\exp[i{\bf k
\delta_{\alpha}}]$ we write this equation  as
\begin{eqnarray}
&&  1   =   \frac{J}{2 \,
N} \sum_{\bf q}
  \sum_{\beta} \exp[ i{\bf q}  (\delta_{\beta} - \delta_{\alpha})] e^{i(2\pi/3)(\beta - \alpha)}
 \nonumber\\
 &\times &
 \Bigl[\frac{ 1}{2 \varepsilon_{+}({\bf q})}
 \tanh \frac{\varepsilon_{+}({\bf q})}{2T_c} + \frac{ 1}{2 \varepsilon_{-}({\bf q})}
 \tanh \frac{\varepsilon_{-}({\bf q})}{2T_c}\Bigr].
  \nonumber\\
    \label{g32b}
\end{eqnarray}
 Considering the vectors $a_{\beta \alpha} \equiv \delta_{\beta} -
\delta_{\alpha}$ we can  see that different values of $\beta - \alpha$ correspond to a
set of  three vectors, one is equal to zero, and the other two are the lattice vectors.
The change of $\alpha$ rotates the whole set by $2\pi/3$ due to the $C_3$ symmetry of the
lattice so that the above equation  does not depend on $\alpha$. Therefore, all three
equations for various $\alpha$ are the same. The equation for $\alpha = 0$ reads:
\begin{eqnarray}
&& 1 = \frac{J}{ N} \sum_{\bf q}
    \frac{1}{2} \sum_{\beta} \exp[i{({\bf q}  ({\bf \delta}_{\beta} - {\bf \delta}_{0})}] e^{i(2\pi/3)\beta}
 \nonumber\\
 &\times &  \Bigl[\frac{ 1}{2 \varepsilon_{+}({\bf q})}
 \tanh \frac{\varepsilon_{+}({\bf q})}{2T_c} + \frac{ 1}{2 \varepsilon_{-}({\bf q})}
 \tanh \frac{\varepsilon_{-}({\bf q})}{2T_c}\Bigr] .
  \nonumber\\
    \label{g32c}
\end{eqnarray}
\begin{figure}
\centering
\resizebox{0.4\textwidth}{!}{%
\includegraphics{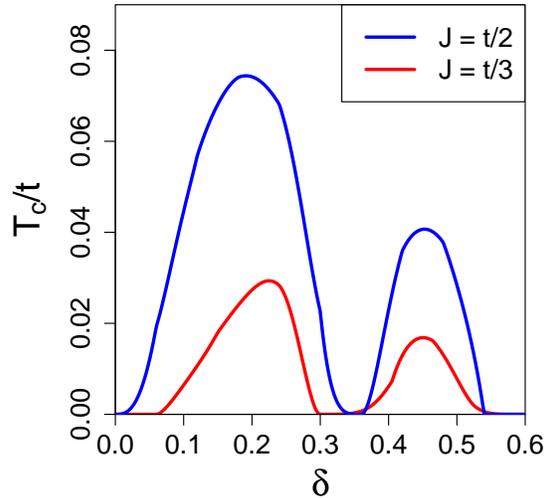}}
\caption{(Color online) $T_c(\delta)$ for the $d$-wave pairing with  $J =  t /3 $ (dotted curve) and $J =
t/2$ (bold line). }
 \label{fig2}
\end{figure}
This equation was solved numerically for various values of $J/t$.  The results for $T_c$
as a function of the hole doping $\delta$ are depicted  in Fig.~\ref{fig2} and
Fig.~\ref{fig3}. $T_c(\delta)$ has a two-peak structure with the maxima corresponding to
the VHS in the DOS shown in  Fig.~\ref{fig1}.

The transition temperature $T_c$ rapidly increases with the interaction $J/t$ in
Eq.~(\ref{g32c}), as was also found in Ref.~\cite{Black07}. For small $J/t = 1/3$ in
Fig.~\ref{fig2} we get a smooth increase of $T_c$  with $\delta $ and  $T_c = 0$ (or
exponentially small) for $\delta < 0.05$ similar to  the  $d$-wave pairing in the
one-band $t$--$J$ model on the square lattice (see Ref.~\cite{Plakida99}). In
Ref.~\cite{Gu13}  for $J/t = 1/3$ the superconducting order at $T =0$  was found for $0 <
\delta < 0.4$,  but in Ref.~\cite{Black07} $T_c$ is exponentially small for $\delta <
0.05$ when $J = 0.8 \, t$.
\begin{figure}
\centering
\resizebox{0.4\textwidth}{!}{%
\includegraphics{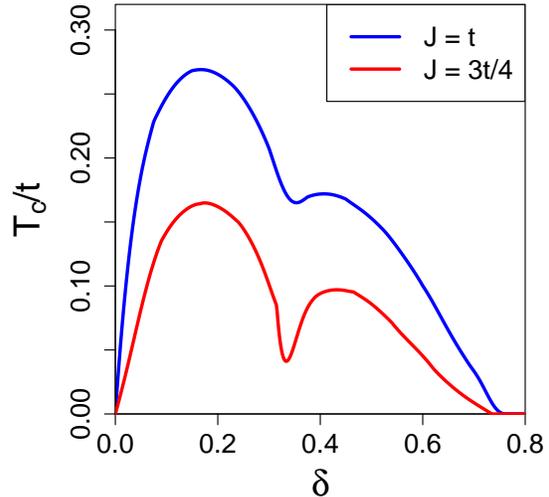}}
\caption{(Color online) $T_c(\delta)$ for the $d$-wave pairing with $J = 3 t /4 $ (dotted curve)
and $J = t $ (bold line). }
 \label{fig3}
\end{figure}
For $J  = 3\,t/4 $ in Fig.~\ref{fig3} the maximum of $T_c \approx  0.15 t$ is comparable
with the maximum value of $T_c \approx 0.1 t$ in Ref.~\cite{Black14} for $J/t = 0.8$ and
larger than $T_c \approx 0.05 t$ in Ref.~\cite{Black07} for $J/t = 0.8$. As was claimed
in these publications, such   values of $T_c$ with the hopping parameter $t \approx
2.5$~eV in graphene would result in a room high-$T_c$ SC of electrons on the honeycomb
lattice at optimal doping. The variational Monte Carlo calculation for the RVB theory in
Ref.~~\cite{Pathak10} shows that   $T_c$ can be estimated as  about twice of the room
temperature.  For small values of $J < J_c = 0.75\,t \, $ there is no pairing at
half-filling for $\tilde{\mu} =0$. For large values of $J/t$ a sharp increase of $T_c$
with $\delta $ is found  in Fig.~\ref{fig3},  and for $J > J_c$  $T_c$  is nonzero at
$\tilde{\mu} =0$. In this region we can observe the gapless SC, as was also found
in Ref.~\cite{Uchoa07} for the $s$-wave pairing in graphene for the nn BCS
coupling parameter $g_1 > g_1^c$.

Taking into account the results of Ref.~\cite{Vladimirov18}  for the $t$--$J$  model on
the honeycomb lattice, where AF long-range order at $T = 0$  was observed for $ \delta
\lesssim 0.1$,  we argue that the AF and superconducting ground-state order may coexist
in the range $ \delta \lesssim 0.1$, in agreement with the numerical calculations in
Ref.~\cite{Gu13}.

Let us compare the results for the extended $s$-wave pairing and $d$-wave pairing. As was
found  in Sect.~\ref{sec:4b},   the maximum value of  $T_c^{\rm max} \approx 0.082\, t
\,$ for   $J =   t/2 $ is nearly the same as in the case of $d$-wave pairing in
Fig.~\ref{fig2}. It contradicts  to the results of Ref.~\cite{Black07}, where $T_c$ for
the $d$-wave pairing is much  larger than for the extended $s$-wave pairing. Note that in
Ref.~\cite{Black07} the mean-field RVB theory was used,  where the restriction of
no-double-occupancy has not been taken into account.   On the other hand, in
Refs.~\cite{Black14,Wu13}, where the renormalized mean-field approximation determined by
the parameters $g_t$ and $g_J$ for the $t$--$J$ model  was employed,  the maximum value
of the order parameter (and $T_c$) for the extended $s$-wave pairing and the  $d$-wave
pairing are quite close. So we can say that accounting for  strong correlations  both in
our  approach and in the renormalized mean-field approximation results in close maximum
values of $T_c$ for the extended $s$-wave  and $d$-wave pairing. However, it should be
stressed that the constraint (\ref{c1}) excludes the $s$-wave pairing. This conclusion is
supported by the calculations for the $t$--$J$  model in Ref.~\cite{Gu13}, where the
projected character of the electron operators has been implemented and only $d$-wave
pairing has been found.

\subsection{Constraint for the s-wave pairing}
\label{sec:c}

Let us consider the restriction of  no-double-occupancy (\ref{c1}) for the Hubbard
operators:
\begin{eqnarray}
\langle    X_{i A}^{0 \bar{\sigma} }X_{iA}^{0 \sigma}\rangle  =  0 .
     \label{c1a}
\end{eqnarray}
Using the GF $F_{AA}^{\sigma}({\bf k}, \omega)$ (\ref{e15}) this condition reads
\begin{eqnarray}
&& \langle    X_{i A}^{0 \bar{\sigma} }X_{iA}^{0 \sigma}\rangle =\frac{1}{N}\, \sum_{\bf
k} \int_{-\infty}^{\infty}\frac{d\omega }{\exp(\omega/T)+ 1}
 \nonumber\\
  \nonumber\\
&& \times
 [- \frac{1}{\pi}]\,{\rm Im}\frac{Q \, A_{31}({\bf k}, \omega)}
{ [\omega^2 -\varepsilon^2_{+}({\bf k}) ] [\omega^2 -\varepsilon^2_{-}({\bf k})  ]} =0 .
          \label{c2}
\end{eqnarray}
where the  function $  A_{31 \sigma}({\bf k}, \omega) $ is given by
(\ref{e15a}). This condition for  the   $s$-wave pairing determined by the gap function
$\Delta_{\sigma}$ (\ref{a8}),  where $ A_{31 \sigma}(\omega =
\varepsilon_{\pm}({\bf k})) =\mp 2 \Delta_{\sigma}\, \widetilde{ \mu }\widetilde{t}\,
  |\gamma ({\bf k})|$,  results in the relation:
\begin{eqnarray}
\langle    X_{i A}^{0 \bar{\sigma} }X_{iA}^{0 \sigma}\rangle &= & - \frac{Q\,
\Delta_{\sigma} }{2 N}\, \sum_{\bf k}  \Bigl[\frac{ 1}{2 \varepsilon_{+}({\bf k})}
 \tanh \frac{\varepsilon_{+}({\bf k})}{2T}
 \nonumber\\
 &&  + \frac{ 1}{2 \varepsilon_{-}({\bf k})}
 \tanh \frac{\varepsilon_{-}({\bf k})}{2T}\Bigr] .
     \label{c3}
\end{eqnarray}
The summation  over $ {\bf k} $  does not vanish for a positively defined integrand.
Therefore,  the  $s$-wave pairing with the ${\bf k}$-independent gap $\Delta_{\sigma} $
violates the kinematic restriction (\ref{c2}) and is ruled out.

For the extended  $s$-wave pairing with the gap (\ref{e17})  we have
$ A_{31 \sigma}(\omega= \varepsilon_{\pm}({\bf k}) ) = 2 \Delta_{AB \sigma} \,\widetilde{
\mu}\,\widetilde{t}\, |\gamma ({\bf k})|^2$, and the relation (\ref{c2}) reads:
\begin{eqnarray}
\langle    X_{i A}^{0 \bar{\sigma} }X_{iA}^{0 \sigma}\rangle &= & - \frac{Q\, \Delta_{AB
\sigma}}{2 N}\, \sum_{\bf k}  |\gamma ({\bf k})| \Bigl[\frac{ 1}{2 \varepsilon_{+}({\bf
k})}
 \tanh \frac{\varepsilon_{+}({\bf k})}{2T}
\nonumber\\
 &&  - \frac{ 1}{2 \varepsilon_{-}({\bf k})}
 \tanh \frac{\varepsilon_{-}({\bf k})}{2T}\Bigr].
     \label{c4}
\end{eqnarray}
The summation  over $ {\bf k} $  does not vanish except for $\widetilde{ \mu} =0$ when
$\varepsilon_{+}({\bf k}) = - \varepsilon_{-}({\bf k})$. For other doping the correlation
function (\ref{c4}) is non-zero that violates the kinematic restriction (\ref{c2}), and
the extended $s$-wave pairing cannot  be realized.

Finally, let us consider the  $d$-wave pairing  with the bond-dependent gap (\ref{e18}).
In this case, for the function   (\ref{e15a}) we have
 \begin{eqnarray}
&& A_{31}(\omega= \varepsilon_{\pm}({\bf k}) )
  = 2\widetilde{ \mu }\, \widetilde{t}\, \gamma ({\bf k})
  \sum_{\alpha}\Delta^{\alpha}_{\sigma}
\exp[-i {\bf k  \delta_{\alpha}}]
\nonumber\\
  &  &  \pm \widetilde{t}^2\, |\gamma ({\bf k})|  \,
\sum_{\alpha}\Delta^{\alpha}_{\sigma} \, 2 i {\rm Im}[\gamma ^*({\bf k}) \exp(i{\bf k
\delta_{\alpha}})],
  \label{c5}
\end{eqnarray}
where the relation $\,\gamma^*({\bf k}) \exp(i{\bf k  \delta_{\alpha}}) - \gamma({\bf k})
\exp(- i{\bf k  \delta_{\alpha}})  =  2 i\, {\rm Im}(\gamma ^*({\bf k}) \exp(i{\bf k
\delta_{\alpha}})) $  was used.  For the correlation function (\ref{c2}) we obtain:
 \begin{eqnarray}
&&  \langle X_{iA}^{0 {\sigma} }  X_{iA}^{0\bar{\sigma}}\rangle
    =  \frac{Q}{2 N}\,\sum_{\alpha}\Delta^{\alpha}_{\sigma} \sum_{\bf k}
   \frac{\gamma ({\bf k})}{|\gamma ({\bf k})|}
 \exp(-i {\bf k } \delta_{\alpha})
\nonumber\\
 & \times & \Big [\frac{ 1}{2 \varepsilon_{+}({\bf k})}
 \tanh \frac{\varepsilon_{+}({\bf k})}{2T} -
 \frac{ 1}{2 \varepsilon_{-}({\bf k})}
  \tanh \frac{\varepsilon_{-}({\bf k})}{2T} \Big]
\nonumber\\
  &+ &  \frac{ Q\, \widetilde{t}}{4 \widetilde{ \mu }}\,  \frac{1 }{N}\, \sum_{\bf k} \
   \sum_{\alpha}\Delta^{\alpha}_{\sigma} \, 2i\, {\rm Im}[\gamma ^*({\bf k})
\exp(i{\bf k  \delta_{\alpha}})]
 \nonumber\\
 & \times  & \Big [\frac{ 1}{2 \varepsilon_{+}({\bf k})}
 \tanh \frac{\varepsilon_{+}({\bf k})}{2T} +
 \frac{ 1}{2 \varepsilon_{-}({\bf k})}   \tanh \frac{\varepsilon_{-}({\bf k})}{2T} \Big].
 \label{c6}
\end{eqnarray}
In the first term the summation  over  ${\bf k }$ does not depend on $\alpha$ due to the
$C_3$ symmetry of $\gamma ({\bf k})$  and $\varepsilon_{\pm}({\bf k})$. Therefore, the
summation over $\alpha$ of the gap function  $\Delta^{\alpha}_{\sigma}$ can be done
independently that results in the vanishing of the first term: $
\sum_{\alpha}\Delta^{\alpha}_{\sigma} = \Delta_{1 \sigma}
\sum_{\alpha}\exp(i(2\pi/3)\alpha)= 0$. The second term also gives no contribution due to
summation over  ${\bf k } $ of the odd in ${\bf k }$ function ${\rm Im}[\gamma ^*({\bf
k})\exp(i{\bf k \delta_{\alpha}})] = i \sum_{ \beta} \sin[{\bf k }( \delta_{\alpha}-
\delta_{\beta}) ] $. So, the $d$-wave pairing  with the gap (\ref{e18}) does not violate
the condition of no-double-occupancy (\ref{c2}). This conclusion  was checked by the
direct integration over ${\bf k}$ in Eq.~(\ref{c6}).

\section{Conclusion}
\label{sec:5}

In this paper a  microscopic theory of superconductivity in  electronic systems with
strong correlations is presented within the $t$--$J$  model on the honeycomb lattice. The
constraint  of no-double-occupancy  in the two-band $t$--$J$  model is rigorously taken
into account by employing the HO technique. The superconducting $T_c$ as a function of
doping is calculated for the $d+id'$ gap function. It reveals a two-peak structure
related to the two VHS in the two-band electronic spectrum. For large values of $J > J_c
= 0.75 t$ a gapless superconductivity is found at $\tilde{\mu} = 0$. It is suggested that
for small doping, $\delta \lesssim 0.1$, the AF long-range order found
in~\cite{Vladimirov18} may coexist with the $d+id'$ superconductivity.

We have also calculated   $T_c$ for  the extended $s$-wave pairing to compare it with the
results obtained for the $t$--$J$  model, where the constraint has been neglected  or
considered approximately. In the latter case the results for the maximum value of  $T_c$
are comparable. However, we have shown that the  $s$-wave pairing violates the constraint
and should be ruled out.

In graphene with the large hopping parameter $t \approx 2.5$~eV the single-site Coulomb
repulsion is not strong enough,  $U/t =4 -5$~\cite{Neto09}. Therefore, the application of
the $t$--$J$  model to graphene is questionable. In our theory we have the renormalized
hopping parameter $\tilde{t}$ (\ref{a6}) which is small  in the region  $\delta \leq
0.4$, where the nn correlation function $\,C_1 \leq  -0.1$~\cite{Vladimirov18}.
Therefore, we have $U/\tilde{t} \gg 1$, and the application of the $t$--$J$  model can be
justified. Moreover, complicated structures like sulfur-graphite
composites~\cite{Black07} may result in a larger value of $U/t$,  and high-$T_c$
superconductivity can be achieved.

\acknowledgments
 The financial support by the Heisenberg-Landau program of JINR is
acknowledged. One of the authors (N. P.) thanks the Directorate of the MPIPKS for the
hospitality extended to him during his stay at the Institute, where a part of
the present work has been done.\\

\end{document}